\magnification 1200

\font\titlefont=cmss17 at 29.88 true pt
\font\authorfont=cmssi17 at 20.74 true pt
\font\bigaddressfont=cmss12 at 14.4 true pt 
 at 14.4 true pt
\def \mysubmit {}
\def \mypresent {}

\def \docnum #1 { \def \mydocnum {#1}} 
\def \date #1 { \def \mydate {#1}} 
\def \title #1 {\def \mytitle {#1}}
\def \author #1 {\def \myauthor {#1}}
\def \abstract #1 {\def \myabstract {#1}}

\def \tobesubmittedto #1 { \def \mysubmit {\leftskip=0pt plus 1fill \rightskip=0pt plus 1fill 
\hbox{\vbox{\noindent \hfill \it To be 
submitted To #1 \hfill \ }}}}
\def \submittedto #1 { \def \mysubmit {\leftskip=0pt plus 1fill \rightskip=0pt plus 1fill 
\hbox{\vbox{\noindent \hfill \it Submitted 
To #1 \hfill \ }}}}
\def \presentedat #1 { \def \mypresent {\leftskip=0pt plus 1fill \rightskip=0pt plus 1fill 
\hbox{\vbox{\noindent \hfill \it Presented 
at #1 \hfill \ }}}}

\def\maketitle{
\let\footnotesize\small
\let\footnoterule\relax

\ifx\mydate\undefined \def \mydate {
\ifcase\month\or
January\or February\or March\or April\or May\or June\or
July\or August\or September\or October\or November\or December\fi
\space\number\day, \number\year} \fi

\ifx\thispagestyle\undefined \nopagenumbers \fi
\ifx\nopagenumbers\undefined {\thispagestyle{empty}} 
      \setcounter{page}{0}%
      \fi
\null
\rightline{\logo}
\vskip 2 pt
\rightline{\mydocnum}
\rightline{\mydate}
\vskip 20 pt
{\def\\{\break} 
\leftskip=0pt plus 1fill
\rightskip = 0 pt plus 1fill
\parindent 0 pt
\baselineskip 30 pt
\titlefont \hfil \vbox { \mytitle}\hfil }
\vskip 25 pt
{\def \and {\qquad} 
\leftskip=0pt plus 1fill
\rightskip = 0 pt plus 1fill
\parindent 0 pt 
\authorfont
\lineskip 12 pt
\myauthor
\parfillskip=0pt\par
}%
\vskip 20 pt

{\bigaddressfont
\centerline{ Department of Physics}
\centerline{ Manchester University}
\centerline{ England}
}

\ifx\myabstract\undefined {}
\else
\null\vfil\vskip 10 pt
\centerline{ \bf Abstract}
\vskip 10 pt
\myabstract
\fi

\ifx\footline\undefined   
\begin{figure}[b]
\mypresent
\mysubmit
\end{figure}
\mythanks
\setcounter{footnote}{0} 
\vfil
\null
\else                       
\footline={{\baselineskip=10 pt \vbox{\hbox to \hsize {\mypresent} \vskip 5 pt \hbox to \hsize{\mysubmit}}}}
\vfil
\eject
\pageno=1 
\footline={\hss\tenrm\folio\hss}
\fi
				      
\let\thanks\relax
\gdef\mysubmit{}\gdef\present{}
\gdef\mythanks{}\gdef\myauthor{}\gdef\@title{}\let\maketitle\relax}

\def \phonenumber{4170}
\def\today{\number\day/\number\month/\number\year\space\number\hour%
:\number\minute\space\jobname}

\def \logo {\vbox to 23.5 mm {
\hbox {}
\hbox to 47 mm 
{
\includegraphics{../tex/logo.ps} 
\hfil} \vfil }}

\def\header #1\par{ \centerline{\it #1}\par}
\def \letterhead {
\voffset -10 mm
{\advance \hsize by 2 cm
\font\address=cmss12
\vskip -9.5 pt
{\address
\vbox {
\vskip 2 mm
\hbox to 6 cm {\hskip -1 cm Department of Physics and Astronomy\hfill }
\hbox to 6 cm {\hskip -1 cm The University of Manchester \hfill }
\hbox to 6 cm {\hskip -1 cm Manchester \hfill }
\hbox to 6 cm {\hskip -1 cm M13 9PL \hfill }
\hbox to 8 cm {\hskip -1 cm Tel 0161-275-\phonenumber\hskip 1.0 cm 
Fax 0161-273-5867\hskip 1.0 cm}
\hfill
\vbox{
\hbox to 6 cm{\includegraphics{../tex/logo.ps}} 
\vskip 0.5 mm
}
}
}}
}

\def \letter #1 {
\topskip 0 pt
\vsize 599 pt
\nopagenumbers
\letterhead
\vskip 1 mm

\vbox to 3.4 cm {\vfill #1 \vfill}
\vskip 5 mm

\hbox to 2 cm {\hskip -1 cm \leaders\hrule height .5 pt \hfill \hskip 1 cm}

\rightline {\number\day \
\ifcase\month\or January\or February\or March\or April\or
May\or June\or July\or August\or September\or October\or
November\or December\fi \ \number\year}

\vskip 10 mm
}

\def \AddressRAL #1 { \leftline{#1 }
\leftline{HEP Division,}\leftline{Rutherford Appleton Laboratory,}
\leftline{Chilton,}\leftline{Didcot,}\leftline{Oxon}}

\def \AddressRegistrar  #1 {
\leftline{#1 }
\leftline{The Registrar's Department}
\leftline{Main Building}
\leftline{University of Manchester}
\leftline{Oxford Road}
\leftline{Manchester M13 9PL}}

\def \AddressFaculty #1 {
\leftline{#1 }
\leftline{The Faculty of Science}
\leftline{Roscoe Building}
\leftline{University of Manchester}
\leftline{Oxford Road}
\leftline{Manchester M13 9PL}}

\def \AddressPhysics #1 {
\leftline{#1 }
\leftline{Department of Physics}
\leftline{University of Manchester}
\leftline{Oxford Road}
\leftline{Manchester M13 9PL}}

\def \AddressCERN #1/#2 {
\leftline {#1}
\leftline {#2 Division,}
\leftline {CERN,}
\leftline {CH1211 Gen\`eve 23,}
\leftline {Switzerland}}

\def \NIM #1 {{\it Nucl. Instr \& Meth. \/}{\bf A#1}\ }
\def \ZPC #1 {{\it Zeit. Phys. \/}{\bf C#1}\ }
\def \NPB #1 {{\it Nucl. Phys. \/}{\bf B#1}\ }
\def \PLB #1 {{\it Phys. Lett. \/}{\bf B#1}\ }
\def \PL #1 {{\it Phys. Lett. \/}{\bf #1}\ }
\def \PRD #1 {{\it Phys. Rev. \/}{\bf D#1}\ }
\def \PRL #1 {{\it Phys. Rev. Lett.\/}{\bf #1}\ }
\def \PR #1 {{\it Phys. Rev. \/}{\bf #1}\ }
\def \CPC #1 {{\it Comp. Phys. Comm. \/}{\bf #1}\ }

\newcount \eqnumber
\newcount \fignumber
\newcount \highref

\def \eq #1{\global\advance \eqnumber by 1
\let \rrr=\eqnumber
\xdef #1{\the\rrr}
\eqno (\the\eqnumber)
}
\def \fig #1{Figure \global\advance \fignumber by 1 \the\fignumber
\let \rrr=\fignumber
\xdef #1{Figure \the\sss}
}

\newcount\refcheck
\newcount\thisrf
\def\references #1 {
\thisrf=0
\ifnum\refcheck=0
\else
\leftline{\bf References}
\fi
\input #1
\refcheck=1
}

\def\refiopstyle #1:#2;#3\par{
\relax
\ifnum\refcheck=0
\edef \rrr{#2}
\let #1=\rrr
\else
#2 
\ 
#3
\par
\fi
\relax
}

\def\refseq#1{\ifnum#1>\the\highref\global\advance\highref 1  
\ifnum#1>\the\highref
\message{Reference out of Sequence: expecting \the\highref got #1}
\highref=#1\fi\fi}

\def\reference #1:#2\par{\advance \thisrf by 1
\relax
\ifnum\refcheck=0
\let \sss=\thisrf
\edef \rrr{\the\thisrf\noexpand\refseq{\the\thisrf}}
\let #1=\rrr
\else
\the\thisrf
:\ 
#2
\par
\fi
\relax
}

\font\small=cmr8

\def \tickbox {\lower 2 mm \hbox{
\vbox  {\vskip .5 mm\hrule \hbox to 4 mm
{\vrule \strut  \hfill  \vrule}\vfill\hrule} }}

\newcount\secnum
\newcount\examplenum
\newcount\subnum
\newcount\subsubnum
\newcount\chapnum
\font \splash=cmssi17
\font \titlefont=cmss17 scaled \magstep2

\font \address=cmss12
\font \cf=cmbxsl10
\examplenum=0
\secnum=0
\subnum=0

\def\chapter #1 {
\advance\chapnum by 1 \secnum=0 \subnum=0 \subsubnum=0
\vfill
\hbox{\bf \quad Chapter \number \chapnum : #1}}
\def\lecture #1 #2{
\secnum=0 \subnum=0
\hbox{\centerline{\splash SLUO Lecture #1: #2}}
\headline={\ifnum\pageno>1 SLUO Lecture #1 \dotfill #2 \fi}
\footline={\ifnum\pageno>1 \hss --\  \folio \ -- \hss  \else  \fi}
}

\def\subsection#1 \par{\par \advance\subnum by 1
\subsubnum=0
\goodbreak \vskip 0.3cm\leftline{\cf \number \secnum .\number 
\subnum \ #1} \par}

\def\subsubsection#1 \par{\par \advance\subsubnum by 1
\goodbreak \vskip 0.3cm\leftline{\sl \number \secnum .\number 
\subnum .\number \subsubnum \ #1} \par}

\def\section#1\par{\goodbreak \par \advance\secnum by 1 \subnum=0
\vskip 0.5cm\leftline{\bf \number \secnum . \ #1}\vskip 0.02cm \par
\message{ Section  \number \secnum    #1}
}

\long\def\example #1 {\par \advance\examplenum by 1
\vskip 12 pt \goodbreak \boxit {{\bf Example \number\examplenum :} #1 }
}

\long\def\boxit #1{\vbox {\kern-5pt\hrule\hbox{\vrule\kern3pt
           \vbox{\kern3pt #1 \kern3pt} \kern3pt \vrule} \hrule}}

\def\bull#1\par {\item {$\bullet$} #1 \par}

\title {A Calculator for Confidence Intervals}
\docnum {MAN/HEP/2001/04}
\author {Roger Barlow} 
\submittedto{Computer Physics Communications}

\docnum {}
\date{}
\def \logo { }

\abstract {A calculator program has been written to give 
confidence intervals on branching ratios for rare 
decay modes (or similar quantities) calculated  from the number of events observed,
the acceptance factor, the background estimate and
the associated errors.
Results from different experiments (or 
different channels from the same experiment) can be combined.
The calculator is available in
{\tt http://www.slac.stanford.edu/\~{}barlow/limits.html}
\hfill\break
\hfill\break
PACS code 13.85.Rm
\hfill\break
\hfill\break
}

\maketitle

\section Introduction

Many particle physics experiments measure a number $n$ of events 
of a particular type,
and use it to set limits on
a physical quantity $R$, typically a cross section or branching ratio.
$R$ and $n$ are related by various factors such as the 
efficiency, dead-time, backgrounds,
beam luminosity, and duration of the experiment.

For a Poisson process, the production of confidence limits on the
mean $\mu$ from the observed $n$ has been well studied  
[1,2]. Tables exist, or standard numerical techniques can be used, to solve
exactly the desired probability equalities 
$$\sum_{r=0}^n P(r;\mu_+)=\alpha \qquad 
\sum_{r=0}^{n-1} P(r;\mu_-)=1-\alpha \eqno(1)$$
for  the upper limit $\mu_+$ and  the lower limit $\mu_-$
 at the $(1-\alpha)$ confidence level.
If the    
factor for efficiency etc., collectively known as the sensitivity $S$, is 
known exactly, then in the absence of background these limits translate directly into
limits on $R$ using 
$$\mu=R S\eqno(2).$$
 If the expected 
background $b$ is known exactly then it can be subtracted
and the confidence limits on $R$ are given by
$$R_\pm= {1 \over S}(\mu_\pm - b)\eqno(3).$$
This note considers cases where 
there are uncertainties 
$\sigma_S$ and $\sigma_b$
in $S$ and $b$. 
It follows the approach of Cousins and Highland [3].
Exact calculation is not possible,
but a Monte Carlo technique can readily be  used.
A trial value for the signal is taken, then repeatedly it has a background
mean drawn from a Gaussian of mean $b$ and standard deviation $\sigma_b$ added
to it, is multiplied by a Sensitivity drawn from a
Gaussian of mean $S$ and standard deviation $\sigma_S$, and is then used as the 
mean of a Poisson distribution from which a number of events $r$
is generated. By counting the
fraction of cases in which $r$ is less than (or equal to)
the observed number $n$, the probabilities $1-\alpha$ (or $\alpha$) of
Equation 1 are given.

\section Philosophy

This is basically a conventional frequentist limit.  There are
no prior assumptions about $R$.
However 
uncertainties in $S$ and $b$
introduce a Bayesian
viewpoint if (as they often do)  they include factors like the 
accuracy of a theoretical model.
Such errors
 are often admittedly not objectively defined:
the experimenter makes statements like `we believe this is good to 10\%'.
 Then the 
statement ``$S=0.50 \pm 0.05$'', if it is taken to
mean that the sensitivity has a 68\% probability of lying
between 0.45 and 0.55, is `unscientific' from a strict frequentist
viewpoint [3]. 

This frequentist/Bayesian admixture is general and inevitable.
It is defensible on the grounds that the main source of
uncertainty -- the Poisson statistics of the number of
events -- is treated in an objective frequentist fashion.
The errors on the factors are small and have small consequences.
The 
ambiguity due to prior distributions are merely
one more uncertainty among many.

\section Results from differing priors

In interpreting a result one naturally writes
$$ R=A n\eqno(4)$$
where 
the `appropriate' factor $A$
is the inverse of the sensitivity $S$.
This was done by the BaBar Statistics Working Group in their report
for the collaboration [5].

It was found that probabilities obtained using the formalism  of 
equation (2) and equation (4), smearing the 
factor  $S$ or $A$ by the same
relative amount, give slightly different results.
Although one is merely the inverse of the other, the relationship
is nontrivial.
A Gaussian  distribution  for $A$ is not 
equivalent to  a Gaussian distribution  for $S$. 
The experimenter may have  an {\it a priori\/} reason for believing that
one or the other has a Gaussian uncertainty,
however this is not the case in general and the choice
is usually merely one of convenience. 

A statistician would say [6] that the
appropriate technique is to  use an `invariant' prior according to
the suggestion of Jeffreys,  i.e. to
work with a uniform prior in the quantity for which the Fisher 
information is constant. For a scale factor $A$ this means a prior
uniform in
 $\ln A$ or, equivalently, $\ln S=-\ln A$.

For example, with no error on the sensitivity the probability of
a signal 5.0 giving a result of 3 events or less is 26.5\%. 
With increasing uncertainty the probability rises, but by slightly different
amounts.

\vskip \baselineskip

\hskip 3 cm \vbox{
\halign{\strut\vrule\ #\hfil &\vrule\hfil \ #\hfil\ &\vrule\ \hfil #\hfil\ &\vrule\ \hfil #\hfil\ \vrule&# \cr
\noalign{\hrule}
 \% error & Cousins \& & Jeffreys & BaBar\cr
on $S$ & Highland & & SWG\cr
\noalign{\hrule}
0 & 0.265 & 0.265 & 0.265 \cr
10\% & 0.272 & 0.269 & 0.266 \cr
20\% & 0.291 & 0.277 & 0.267 \cr
30\% & 0.316 & 0.291 & 0.273 \cr
\noalign{\hrule}
}}
\hfill

\vskip \baselineskip

\centerline{Table 1: Probability for $n\leq 3$ with increasing
uncertainty in the acceptance}

\vskip \baselineskip

Conversely, with 30\% uncertainty the limit of 5.00 gives a
29.1\% probability with a Jeffreys prior; to get the same probability
with the Cousins and Highland hypothesis the limit has to be raised to
5.22, whereas for the BaBar SWG hypothesis it falls to 4.86.  These differences
are not large (and 30\% is somewhat extreme) but this ambiguity does 
affect values at the level of precision with which they are generally
quoted, so it cannot be ignored.
All three values are given by the program, so the user can choose which to use
(indeed, they are forced to be aware of the choice).

\section Combining results

When different experiments, or
different channels in the same experiment, give limits, it is desirable
to combine them.
There is a possible ambiguity in the procedure.  For, say, an upper limit
from a single experiment
one determines a value $R_+$ such that the probability of
observing $n$ events or less is equal to the desired (small) level, as in
Equation 1.
But for a double channel, with results $(n_1, n_2)$, the expression `less'
is undefined: is $(4,1)$ less than $(3,3)$ ?

It could be suggested that `less than' for a pair 
signify that both values should be less, or either
value.  However  these are clearly unworkable:
if the two experiments are identical (in luminosity, efficiency, and
background) then there is clearly no information obtainable from the
distribution between the two: $n_1+n_2$ is a sufficient statistic
for $(n_1,n_2)$, and `less than' can be defined with reference to the
sum of the two results.  Any definition of `less than' must satisfy this.

We propose that `less than' be defined for a multiple result  using 
the value of the Branching ratio that one would deduce from it.
This is given by a maximum likelihood estimator as ([7], Equation 129)  
$$\sum_i {n_i S_i \over S_i R + b_i}-S_i = 0 \eqno(5)$$
where $S_i$ is the sensitivity factor 
for experiment $i$, and $b_i$
is the expected background.
Sets of results can be ranked according to the value of $R$ given by
this equation.
Thus to run the simulation for several experiments one 
can first solve Equation (5) iteratively to find
 $R_{data}$ for the data values,
and then for a given limit $R_+$ repeatedly 
generate random sets of result values
$\{n_i\}$, calculate
the resulting $R$, and count the fraction of cases
that $R\leq R_{data}$.
In practice this can be achieved without an iterative solution for
each result, as  
the quantity in 
Equation 5 is a monotonically decreasing function of $R$,
so having found $R_{data}$  then
for each set of results 
one merely calculates the quantity
$$Q=\sum_i {n_i S_i \over S_i R_{data} + b_i}-S_i \eqno(6) $$
and the sign of $Q$ gives the sign of $R-R_{data}$.

This goes over to the conventional case if all experiments 
have the same $S_i$ and $b_i$.

\section Usage

The user is initally presented with a screen for a single experiment (Figure 1.)
The number
of observed events is entered, as are  the expected
background and its error  (which,  like all errors, is given
as an absolute value),
the overall sensitivity $S$ with its error, 
the number of Monte
Carlo events to be used in the evaluation, 
and a guess for the limit.  
A button is then pressed to run the Monte Carlo simulation, and evaluate
the probability for either an upper or a lower limit.  (The difference is
merely that one includes the $r=n$ result in the probability and
the other does not, according to the two parts of Equation 1.)

\vbox to 6 cm {
{\ }
\vfill
\includegraphics{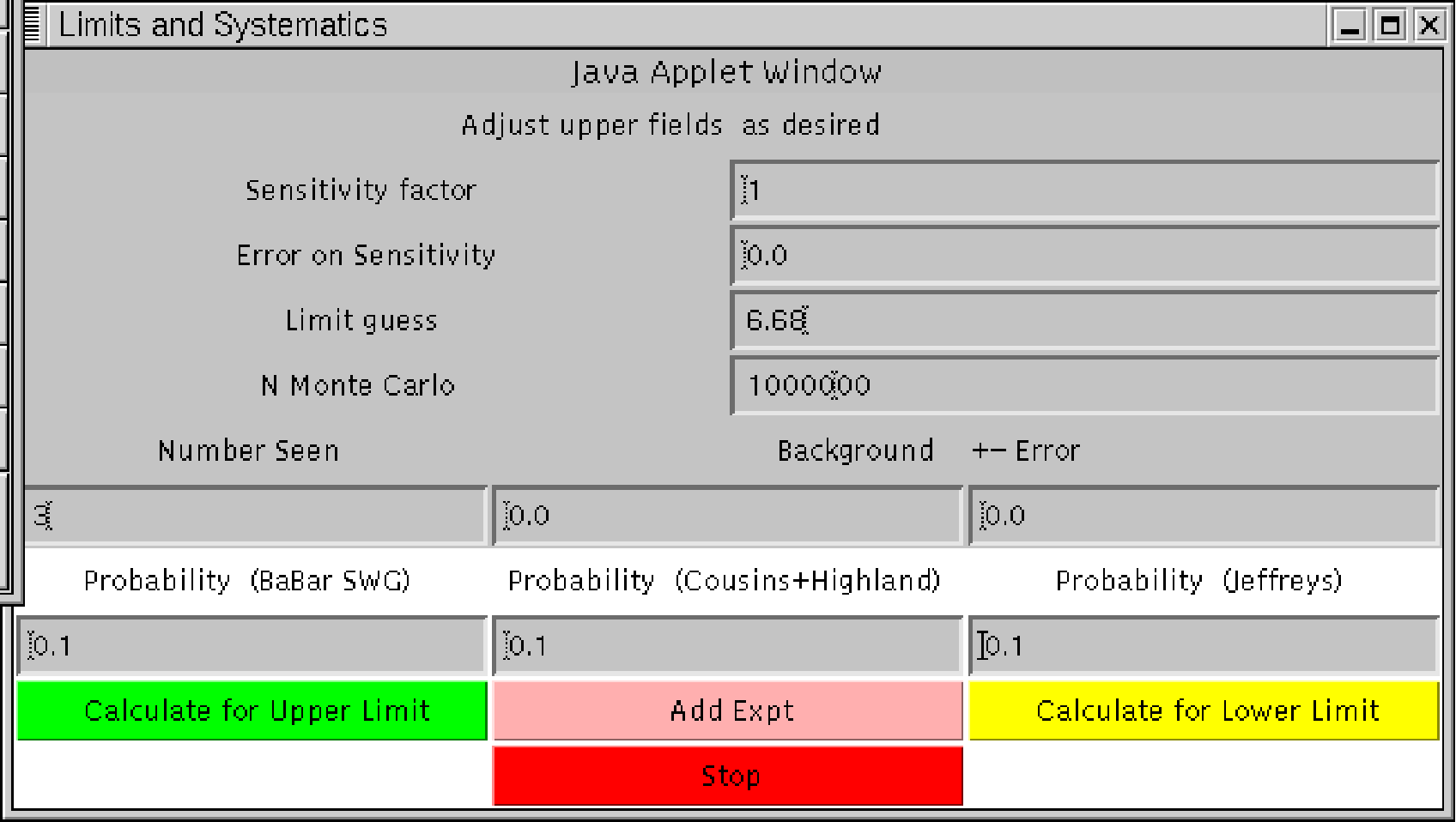}
\centerline{Figure 1: Example of use with a single result}
}

The probability for this limit, under
the three different assumptions about priors, is displayed.
The user can then iterate towards whatever limit type they choose, e.g. for
a 90\% upper limit they adjust the limit guess until the probability
is shown as 0.10; for a 90\% lower limit they would aim for 0.90.   
Only the `Limit guess' data entry field need be  modified for each trial,
although as the limit is approached greater accuracy may be desired
and the number of Monte Carlo events increased accordingly.  (The calculation
takes a few seconds for 1,000,000 events).
In the example shown, 3 events were 
seen with zero background and no sensitivity 
uncertainty.  With 10,000 Monte Carlo events the limit is quickly
established at around 6.7. Increasing the number to 1,000,000 
establishes it as 6.68 (which could have been obtained from tables
anyway). 

It is helpful to remember that the Poisson mean used is the product of the
limit guess and the (smeared) sensitivity factor.
It  may be convenient to work in 
`event-count' units, taking the sensitivity as unity (
with an error which is the the proportional error).  The resulting limit is
equivalent to a number of events and the sensitivity factor 
is applied afterwards:
 if in the above example the data sample 
corresponds to $20 fb^{-1}$ and the efficiency
is 100\%  this 
establishes the limit as $6.68\div 20 = 0.334 fb$. Alternatively the Sensitivity factor
could have been entered directly as 20, and the limit guess found 
to be 0.334.

If the efficiency were, say,  
$50 \pm 5 \%$
this could be accommodated either by setting the sensitivity error to
0.1 and dividing the resulting limit 
(risen to 6.81, using the
Cousins and Highland prior)
by a further 0.5,  
to give $0.681 fb$, or by changing the sensitivity to 10 with an error of 1
and extracting a limit of 0.681 directly.

If the background is estimated as $0.5 \pm 0 $ then the limit falls from
6.81 to to
6.31. For a background of $0.5 \pm 0.2$ it falls to 6.29.

If further experiments are desired (up to a limit of 10) there
is a button to add data fields for them.
There is now also a specific factor and
error for 
each experiment, and this is entered in a further two columns,
as shown in Figure 2.

\vbox to 6 cm {
{\ }
\vfill
\includegraphics{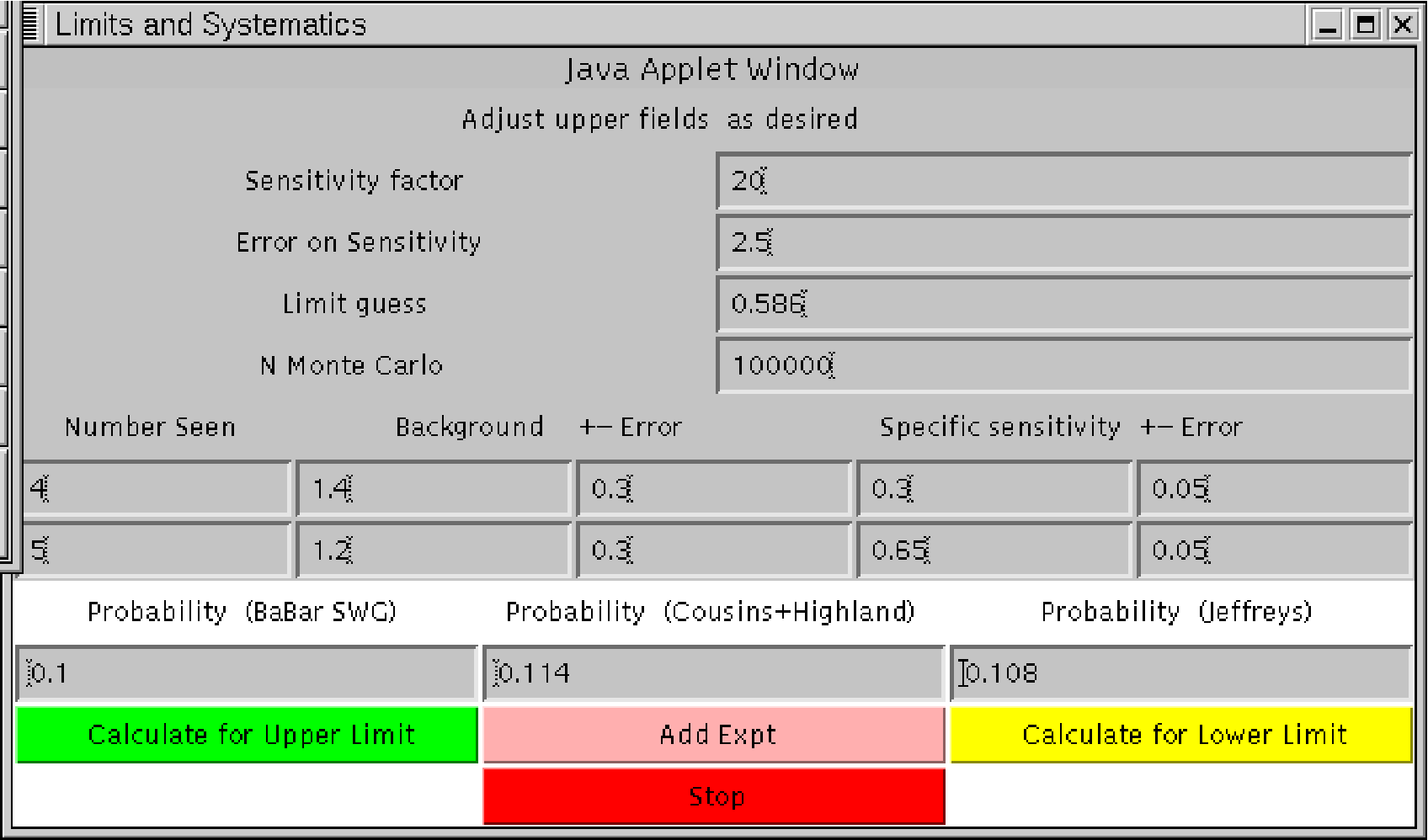}
\centerline {Figure 2: Example of use with two results}
}

For multiple results `event count' units are not
workable.  The sensitivity is split into an overall
value and a specific factor  for each
channel: 
the Poisson mean for an experiment is the product of the
three values entered for the limit, the smeared overall sensitivity, and the 
smeared specific sensitivity. 

The ambiguity in the factors (if the
overall sensitivity is scaled up and each channel's fraction scaled down
by the same amount this makes no difference) is there for convenience
in the handling of errors. 
The common acceptance is varied together for all channels, while the 
specific factors vary independently.
For example, if a single experiment is measuring the rate of $K^0_s$ 
production, using both the $\pi^+ \pi^-$ and $\pi^0 \pi^0$ decay
modes, then the overall luminosity figure and global trigger efficiency
(with associated errors) form the common value.  The two specific factors
are formed from the two branching ratios and the individual reconstruction
efficiencies for making $K^0$ particles from  tracks and from clusters.
On the other hand, if two separate experiments were measuring the rate of
$Z$ production through a decay mode to $\mu^+\mu^-$, 
the common acceptance would
contain the  branching ratio $Z \to \mu^+ \mu^-$ (and associated error)
and the specific factors contain the luminosities etc of the two experiments.

\section Application to a related problem

A different 
problem 
arises where an overall limit is required from two (or more)
channels for which the efficiencies are known, but the
branching ratio is not.  For example, suppose the $X$ particle is 
hypothesised to decay to
$\mu^+\mu^-$ and  $e^+e^-$ in an unknown ratio, but with (different)
efficiencies for detecting $X\to\mu^+\mu^-$ and $X \to e^+e^-$ which are credibly
estimated from Monte Carlo.  From the observed
numbers $N_\mu$ and $N_e$ what can be said about the upper limit
for $X$ production?

This has been  surprisingly difficult to answer.  However we can now do so, at
least in principle.
There are two parameters: $N_X$, the number of $X$ particles, and $p$, the
probability of decaying into muons; the probability of decaying into 
electrons is $1-p$.   We use the fact that in the full
definition of confidence levels, the two parts of Equation 1 are 
inequalities: for the upper limit $N_+$ on $N_X$ the
probability of obtaining this result, or worse, is $\alpha$, {\it or less\/},
 whatever
the value of $p$.  For a given $N_X$ the probability must be maximised by
varying $p$,
$$P(N_X)=Max(P(N_X,p))$$
 and then $N_X$ can be varied until $P(N_X)=\alpha$.
It may be that there is a simple way of finding the value of $p$ which maximises $P(N_X,p)$,
though this is complicated as when $p$ changes not only does the probability for each
$(n_1,n_2)$ value change, but the region which counts as `less than' the
data result changes discontinuously.  However in default of
a solution, the limit can be found using this calculator.

\section Possible future developments

It would 
be possible to relieve the user of the need to guess limits by doing the
iteration within the program. This would require a more complicated front
interface (specifying the confidence level required and the 
interval type: upper, lower, central, shortest-distance...) and an
automatic scanning strategy to locate the limit to some desired accuracy. 

The calculator does not consider the problems that arise when the signal $n$ is 
similar to, or smaller than, the background $b$, i.e. when there has
manifestly been a fluctuation downwards in the background process. This
cannot be taken into account in the standard frequentist framework. 
For such cases extension of the calculation to use either a flat-prior
Bayesian formula 
formula  [8]
or the frequentist Feldman-Cousins technique [9] would  be possible.

The expected background contributions are  specified in terms of 
numbers of observed events. 
The model assumes that the error on this quantity is independent
of the errors on acceptance.  This 
corresponds to cases where backgrounds are estimated, for example, 
from observed sidebands.  
If one has a background from a known physics process
affected by a factor in common with the signal
(e.g. background from Drell-Yan
muon pairs and a serious  uncertainty in the  muon detection efficiency)
then the model could be adapted to do this.
\section References

[1] E. L. Crow and R. S. Gardner, Biometrika {\bf 46} 441 (1959)

[2] A. Stuart and J. K. Ord,  {\it Kendall's Advanced Theory of Statistics\/}
5th Edition, Volume 2, Oxford University Press (1991)

[3] R. D. Cousins and V. L. Highland, \NIM 320 331 (1992)

[4] R. von Mises, {\it Probability, Statistics and Truth\/}, Dover Publications (1981)

[5] The Babar Statistics Working Group,  {\it Recommended Statistical Procedures for Babar}, BaBar Analysis 
Document \# 318. (Version of October 18, 2001)\hfill\break
{\tt http://www.slac.stanford.edu/BFROOT/www/Statistics/Report/report.ps}

[6] 
D. R. Cox and D. V. Hinkley {\it Theoretical Statistics\/}, p 378,
Chapman and Hall, 

\qquad London (1974)

\qquad
D. S. Sivia {\it Data Analysis: a Bayesian Tutorial\/}, p 113,
Clarendon Press,

\qquad  Oxford(1996)

[7] G. Cowan, Journal of Physics G: Nucl. Part. Phys. {\bf 27}, p 1375 (2001)

[8] O. Helene, \NIM 212 319 (1983)

[9] G. J. Feldman and R. D. Cousins, \PRD 57 3873 (1998)

\bye